\documentclass[12pt]{article}
\pdfoutput=1
\usepackage{authblk}
\usepackage[hmargin=1in,vmargin=1in]{geometry}
\usepackage{graphicx}
\usepackage{setspace}
\usepackage{amssymb,amsmath}
\usepackage{xfrac,color,gensymb}
\usepackage[superscript,biblabel]{cite}
\usepackage{pdfpages}
\begin{document}

\renewcommand\thesection{}
\renewcommand\thesubsection{\thesection\arabic{subsection}}
\newcommand{\bra}[1]{\langle #1|}
\newcommand{\ket}[1]{|#1\rangle}
\newcommand{\braket}[2]{\langle #1|#2\rangle}
\newcommand{\brag}{\langle\langle}
\newcommand{\ketg}{\rangle\rangle}
\newcommand{\s}{\sigma}
\newcommand{\clebsch}[3]{\langle #1;#2 \vert #3\rangle}
\newcommand{\expect}[1]{\langle #1 \rangle}
\newcommand{\e}{\varepsilon}
\newcommand{\w}{\omega}
\newcommand{\G}{\Gamma}
\newcommand{\up}{\uparrow}
\newcommand{\down}{\downarrow}
\newcommand{\norm}[1]{\left\lVert#1\right\rVert}
\newcommand{\beq}{\begin{eqnarray}}
\newcommand{\eeq}{\end{eqnarray}}
\newcommand{\be}{\begin{equation}}
\newcommand{\ee}{\end{equation}}
\newcommand{\de}{{\rm d }}
\newcommand{\dd}{\partial}
\newcommand{\h}{\hslash}
\newcommand{\vsd}{V_{\textrm{sd}}}
\newcommand{\ham}[1]{H_{\textrm{#1}}}
\newcommand{\vgate}[1]{$V_{\textrm{#1}}$}
\newcommand{\ts}{$T^*$}
\newcommand{\tc}{$T_c$}
\newcommand{\ybco}{YBa$_2$Cu$_3$O$_{6+\delta}$}
\newcommand{\bi}{(Bi,Pb)$_2$(Sr,La)$_2$CuO$_{6+\delta}$}
\newcommand{\cecusi}{CeCu$_2$Si$_2$}
\newcommand{\cecuau}{CeCu$_{6-x}$Au$_{x}$}
\newcommand{\ybrhsi}{YbRh$_2$Si$_2$}
\newcommand{\etal}{\textit{et al.}}
\newcommand{\insitu}{\textit{in situ}}
\newcommand{\AlxGaAs}[2]{\mbox{$\text{Al}_{#1}\text{Ga}_{#2}\text{As}$}}
\newcommand{\tmatrix}{$\mathcal{T}$-matrix}
\newcommand{\smatrix}{$S$-matrix}

\newcommand{\Startsubfig}[2]{Figure~\ref{fig:#1}#2}
\newcommand{\subfig}[2]{Fig.~\ref{fig:#1}#2}
\newcommand{\allfig}[1]{Fig.~\ref{fig:#1}}
\newcommand{\Startallfig}[1]{Figure~\ref{fig:#1}}
\newcommand{\refeq}[1]{Eq.~\ref{eq:#1}}

\newcommand{\didv}{\mbox{$\text{d}I/\text{d}V_{\rm ds}$}}
\newcommand{\comment}[1]{ {\bf \color{blue} #1} }

\title{\Large \bf Universal Fermi liquid crossover and\\quantum criticality in a mesoscopic system}
\author[1]{\normalsize A.~J.~Keller}
\author[1]{L.~Peeters}
\author[2,3]{C.~P.~Moca}
\author[4]{I.~Weymann}
\author[5]{D.~Mahalu}
\author[5]{V.~Umansky}
\author[2]{G.~Zar\'{a}nd}
\author[1,*]{D.~Goldhaber-Gordon}
\affil[1]{\footnotesize Geballe Laboratory for Advanced Materials, Stanford University, Stanford, CA 94305, USA}
\affil[2]{BME-MTA Exotic Quantum Phases ``Lend\"{u}let'' Group, Institute of Physics, Budapest University of Technology and Economics, H-1521 Budapest, Hungary}
\affil[3]{Department of Physics, University of Oradea, 410087, Romania}
\affil[4]{Faculty of Physics, Adam Mickiewicz University, Pozna\'n, Poland}
\affil[5]{Department of Condensed Matter Physics, Weizmann Institute of Science, Rehovot 96100, Israel}
\affil[*]{Corresponding author; goldhaber-gordon@stanford.edu}
\date{}
\maketitle


{\bf Quantum critical systems derive their finite temperature properties from the influence of a zero temperature quantum phase transition \cite{SachdevBook}. The paradigm is essential for understanding unconventional high-\tc{} superconductors and the non-Fermi liquid properties of heavy fermion compounds. However, the microscopic origins of quantum phase transitions in complex materials are often debated. Here we demonstrate experimentally, with support from numerical renormalization group calculations, a universal crossover from quantum critical non-Fermi liquid behavior to distinct Fermi liquid ground states in a highly controllable quantum dot device. Our device realizes the non-Fermi liquid two-channel Kondo state \cite{Oreg2003,Potok2007:2CK}, based on a spin-1/2 impurity exchange-coupled equally to two independent electronic reservoirs \cite{Nozieres1980:2CK}. Arbitrarily small detuning of the exchange couplings results in conventional screening of the spin by the more strongly coupled channel for energies below a Fermi liquid scale \ts. We extract a quadratic dependence of \ts{} on gate voltage close to criticality and validate an asymptotically exact description of the universal crossover between strongly correlated non-Fermi liquid and Fermi liquid states~\cite{Sela2011,Mitchell2012:Universal}.
}

A conventional second-order quantum phase transition (QPT) features quantum mechanical fluctuations of a classical order parameter. Some second-order QPTs in heavy fermion materials, notably \cecuau{} and \ybrhsi{}, defy easy description in this scheme, and their quantum critical behavior instead appears to be related to the breakdown of Kondo screening.~\cite{GegenwartSiSteglich2008} Distinctive non-Fermi liquid behaviors appear above a so-called Fermi liquid (FL) scale that vanishes at the quantum critical point (QCP); away from the QCP, a crossover from non-FL to FL behavior is observed at low energies. A diverging effective mass $m^*$ at the QCP, seen in both materials, signifies the absence of quasiparticles at the Fermi surface.~\cite{Coleman2001}

In many heavy fermion materials and in high-$T_c$ superconductors, the relevant degrees of freedom and the effective Hamiltonian can be controversial. We aim to understand quantitatively a second-order QPT outside the usual order-parameter-fluctuation description. Quantum dots provide an experimental framework for realizing known quantum impurity Hamiltonians that can feature tunable second-order QPTs \cite{Mebrahtu2012, Mebrahtu2013:Majorana}. However, QCPs are challenging to reach even in engineered systems, since perturbations that steer away from quantum criticality may be inherently uncontrolled, as in two-impurity Kondo experiments to date \cite{Jeong2001:QDMolecule,
Bork2011:2IK,Chorley2012:2IK}.
 
At the QCP of a two-channel Kondo (2CK) system, a single overscreened spin yields a non-FL state with no quasiparticles (i.e. only collective excitations) at the Fermi surface. An order parameter is typically not invoked; rather, the critical behavior is owing to the single spin. A FL scale $T^*$ results from several relevant perturbations: Zeeman splitting, difference in exchange couplings, and charge transfer between the two channels. Requiring that all these perturbations be small would seem to diminish prospects for observing the QCP in bulk systems. Nonetheless, two-channel Kondo physics has been invoked to explain experiments on heavy fermion materials\cite{Cox:HeavyFermion2CK,Seaman:HeavyFermion2CK,Besnus:HeavyFermion2CK} and two-level tunneling centers~\cite{Ralph1994:2nd2CK, Cichorek2005:2CK, Yeh2009:2CK}. A 2CK state has been predicted~\cite{Oreg2003} and observed~\cite{Potok2007:2CK} in a quantum dot tunnel-coupled to a ``metallic grain,'' an electron reservoir big enough to have a small level spacing $\Delta \lesssim kT$ but small enough to retain a charging energy $E_C \gg kT$, at temperatures of interest. The metallic grain provides an independent screening channel, as the grain's charging energy strongly suppresses inter-channel charge transfer. Non-FL behavior was observed, as were the FL single-channel Kondo states far from the QCP, but the crossover to those FL states was not explored. The universal crossover functions were however calculated by NRG~\cite{Toth2007}. Recently, a description of the crossover has been found using Abelian bosonization and conformal field theoretical methods, yielding asymptotically exact predictions for conductance in the regime where $V,T,T^* \ll T_K$~\cite{Sela2011,Mitchell2012:Universal}.

\begin{figure}[hp]
\begin{center}
\includegraphics[width=6.5in]{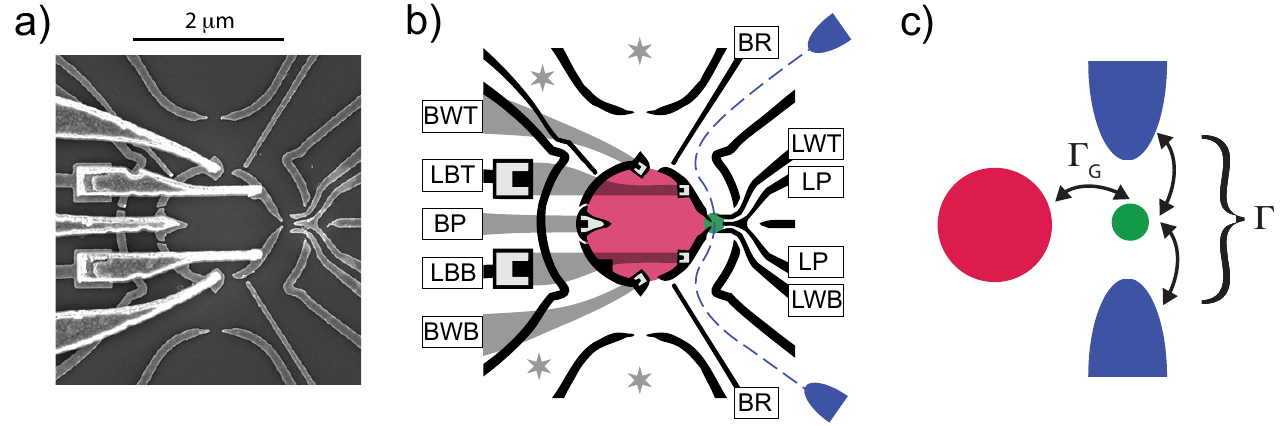}
\end{center}
\caption{\small {\bf Device and model.}
{\bf (a)} SEM micrograph of a nominally identical device. The five brighter features seen coming in from the left are metal bridges suspended above the sample surface.
{\bf (b)} Schematic of the device with labeled gate electrodes. Gates BWT, BP, and BWB define the grain (red) along with LBT and LBB; the last two also control the dot-grain coupling. Gates LWT, LP, and LWB define the dot (green), along with LBT and LBB. Gates BR are used to isolate the dot measurement circuit. Other gates are held at a fixed voltage throughout the experiment. Conductance is measured between source and drain leads (blue). The four gray stars indicate additional ohmic contacts which are floated during measurement.
{\bf (c)} Model of the system used for the NRG calculations. $\Gamma_g$ is the dot-grain coupling, $\Gamma$ the total dot-lead coupling (sum of couplings to source and drain leads). The source and drain leads together act as one channel in the spin 2CK regime, and the Coulomb-blockaded grain acts as an independent channel. Full Hamiltonian in supp. info.
\label{fig:device}
}
\end{figure}

In this work, we show how fine control over the 2CK state in a mesoscopic device allows direct comparison to exact results in the crossover regime, yielding $T^*$ as a function of gate detuning away from the QCP. The device (\subfig{device}{a}) is fabricated by lithographically patterning gate electrodes on a GaAs/\AlxGaAs{0.3}{0.7} heterostructure hosting a two-dimensional electron gas. The device is abstracted in \subfig{device}{b}. Despite the number of gates, the device is conceptually simple (\subfig{device}{c}): a metallic grain (red) and two leads (blue) are each tunnel-coupled to a quantum dot (green) at rates $\Gamma_G$ and $\Gamma$, respectively. The charging energy is $U$ ($E_C$) for the dot (grain) (full Hamiltonian in supp. info). In this experiment, two-terminal conductance $G = dI/d\vsd$ is measured between the pair of leads (Methods sec. \ref{sec:meas}). We use \vgate{BWT} (\vgate{LP}) to tune the grain level $\phi$ (dot level $\varepsilon$).

\begin{figure}[p]
\begin{center}
\includegraphics[width=3.25in]{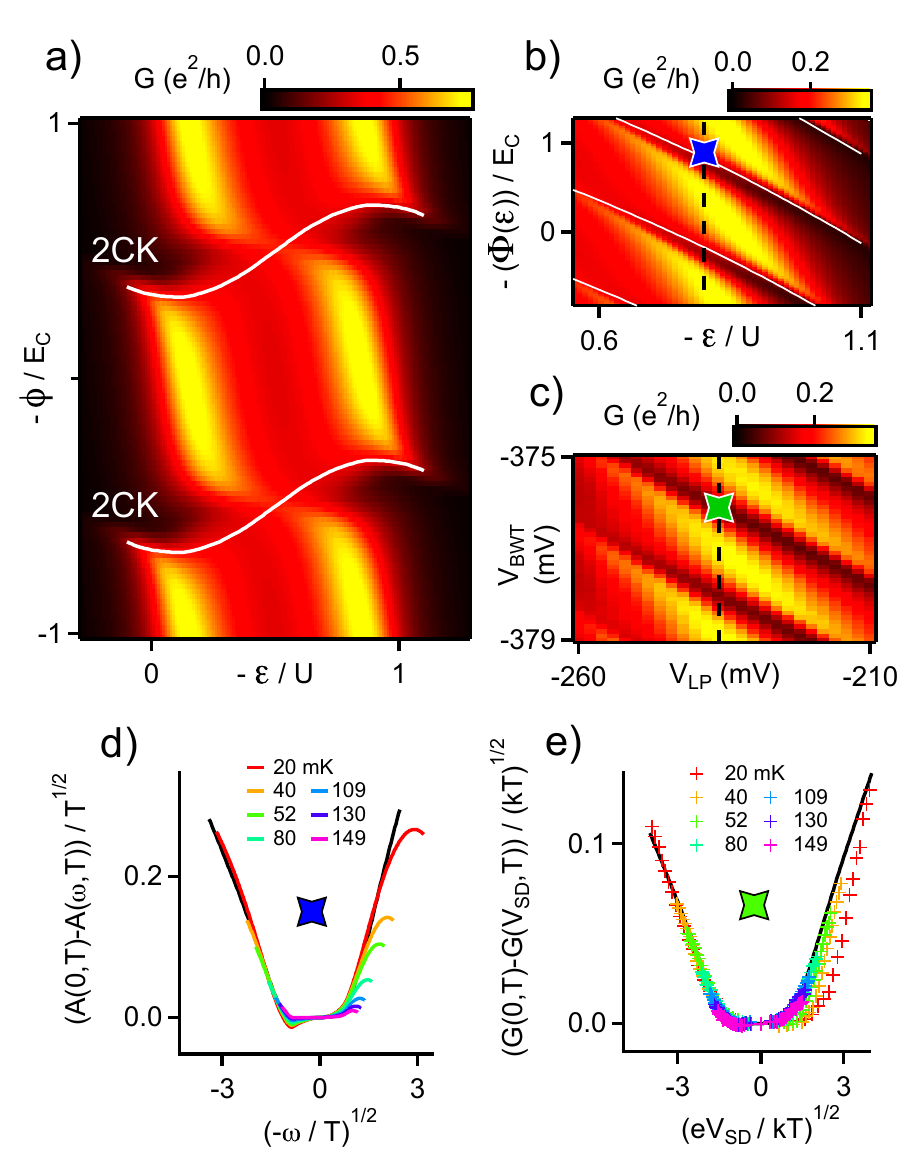}
\end{center}
\caption{\small {\bf Quantum phase transitions.}
{\bf a)} NRG calculations of $G(\vsd=0)$ for symmetric source-drain coupling ($T = 20$~mK). Parameters: $U = 2$~meV,  $\Gamma=0.123$~meV, $\Gamma_G=0.106$~meV, $E_C = 0.15$~meV, bandwidth $D = 1$~meV. 2CK lines are determined by analysis of the finite size spectrum. 
{\bf b)} The calculations of a) plotted with an $\varepsilon$-dependent shift in $\phi$ and rescaled by a constant factor for comparison with c), to account for unequal source-drain couplings. 
White lines indicate 2CK lines.
{\bf c)} Experimentally measured $G(\vsd=0)$. Gates \vgate{BWT} and \vgate{LP} act approximately like $-\Phi$ and $-\varepsilon$. The dashed line indicates the cut direction of \subfig{crossover}{d}.
{\bf d)} NRG calculations of the equilibrium spectral functions $A(\omega,T)$ for $\varepsilon$, $\Phi$ as marked in b). The black trace is the spectral function $A_{\rm{2CK}}(\omega,T,\delta_P)$ from CFT ($\delta_P = -0.029\pi$, $T_K = 19$~$\mu$eV).
{\bf e)} Measured $G(\vsd,T)$ for \vgate{LP}, \vgate{BWT} as marked in c). The black trace is $Y_{\rm{2CK}}(\omega/T,\delta_P)/\sqrt{T_K}$, rescaled based on an estimate of source-drain coupling asymmetry ($\delta_P = -0.016\pi$, $T_K = 50$~$\mu$eV). The range in $(e\vsd/kT)$ decreases as temperature increases because we measure a fixed range in $\vsd$.
\label{fig:qptlines}
}
\end{figure}

We first identify the set of QCPs in the $(-\varepsilon/U, -\phi/E_C)$ plane for fixed $\Gamma, \Gamma_G$. For our model Hamiltonian, quantum critical ``2CK lines'' periodic in the grain charge are expected instead of isolated QCPs \cite{Oreg2003, Anders2004, Anders2006:Proc}. \Startsubfig{qptlines}{a} shows the 2CK lines overlaid on numerical renormalization group (NRG) calculations of $G(-\varepsilon/U, -\phi/E_C)$ using realistic device parameters. We focus on the spin 2CK regime, though charge fluctuations may be important elsewhere \cite{Lebanon2003}. To directly compare to the experimentally measured conductance data of \subfig{qptlines}{c}, \subfig{qptlines}{b} adjusts the NRG calculations of \subfig{qptlines}{a} to account for the cross-capacitance between \vgate{LP} and the grain.

To identify transport signatures of quantum criticality along the 2CK line, we look for the characteristic square-root scaling of $G(\vsd,T)$ derived from the CFT of Affleck and Ludwig \cite{Affleck1993:ExactCFT}. The CFT yields temperature-dependent spectral functions $A_{\rm{2CK}}(\omega,T,\delta_P)$, where $\delta_P$ is a phase shift from potential scattering. These are closely related to $G(\vsd,T)$ for $\omega \rightarrow -e\vsd$ (Methods sec.~\ref{sec:expG}). A scaling collapse of $G(\vsd,T)$ is expected:

\begin{equation}
\frac{G(0,T)-G(\vsd,T)}{\sqrt{T}} \propto \frac{1}{\sqrt{T_K}} Y_{\textrm{2CK}}(-e\vsd/kT, \delta_P)
\label{eq:2ck}
\end{equation}
where $T_K$ is a scale below which the 2CK physics is observed and $Y_{\textrm{2CK}}(-e\vsd/kT,\delta_P)$ a universal function closely related to $A_{\rm{2CK}}(\omega,T,\delta_P)$ (Methods sec.~\ref{sec:2ck}). 

\Startsubfig{qptlines}{d} shows spectral functions $A(\omega,T)$ calculated by NRG. Importantly, the spectral functions collapse onto $A_{\rm{2CK}}(\omega,T,\delta_P)$, with the horizontal axis scaled to emphasize the $\omega^{1/2}$ behavior for large $\omega/T$. Measured $G(\vsd,T)$ on or very near the 2CK line (\subfig{qptlines}{e}) collapse similarly, except for the 20~and~40~mK traces at positive $\vsd$. This deviation could result from a small $T^* \lesssim T_e$, the base electron temperature. Data taken at more negative \vgate{LP} show very clear 2CK scaling (supplemental info) but are less suitable for analyzing the crossover. Experimental $T_K \sim 50$~$\mu$eV should only be trusted up to factors of order unity: in \refeq{2ck}, $T_K$ enters only as a scale factor, and other scale factors like source-drain coupling asymmetry must be estimated.

Having identified the 2CK lines in \allfig{qptlines}, we consider how to perturb the quantum critical state. In the 2CK model, a single FL scale \ts{} suffices to describe any combination of symmetry-breaking perturbations \cite{Sela2011}. The limit $\omega,T,T^* \ll T_K$ permits an exact expression for the scattering $\mathcal{T}$-matrix in the low-temperature 2CK crossover, found by Sela, Mitchell, and Fritz \cite{Sela2011, Mitchell2012:Universal}. In our experimental configuration the $\mathcal{T}$-matrix is diagonal:

\begin{equation}
2\pi i \nu \mathcal{T}_{\sigma \alpha, \sigma \alpha} (\omega,T,\delta_P) = 1 - e^{2i\delta_P} S_{\sigma \alpha, \sigma \alpha} \mathcal{G}\left(\frac{\omega}{T^*},\frac{T}{T^*}\right)
\label{eq:tmat}
\end{equation}
with the universal complex-valued function $\mathcal{G}\left(\frac{\omega}{T^*},\frac{T}{T^*}\right)$ encoding the crossover physics. These diagonal elements relate to $A(\omega,T)$ and thus to experimental $G = dI/d\vsd$ for highly asymmetric source-drain coupling. $\nu$ is the bare density of states per spin in the leads, $\sigma$ is the spin index, and $\alpha=1$ (-1) labels electrons in the leads (grain).  The $S$-matrix gives a (spin and channel dependent) scattering phase shift that is a function of the relative strengths of any perturbations present. Negligible charge transfer between channels and zero magnetic field yields $S_{\sigma\alpha,\sigma\alpha} = \pm \alpha$, with $+ (-)$ indicating the dot is more strongly exchange-coupled to the grain (leads). The factor $e^{2i\delta_P}$ accounts for additional spin-independent phase shifts from potential scattering. We fix $S_{\sigma\alpha,\sigma\alpha} = \alpha$ and let $\delta_P$ jump by $\pi/2$ to account for sign changes.

\begin{figure}[p]
\includegraphics[width=6.5in]{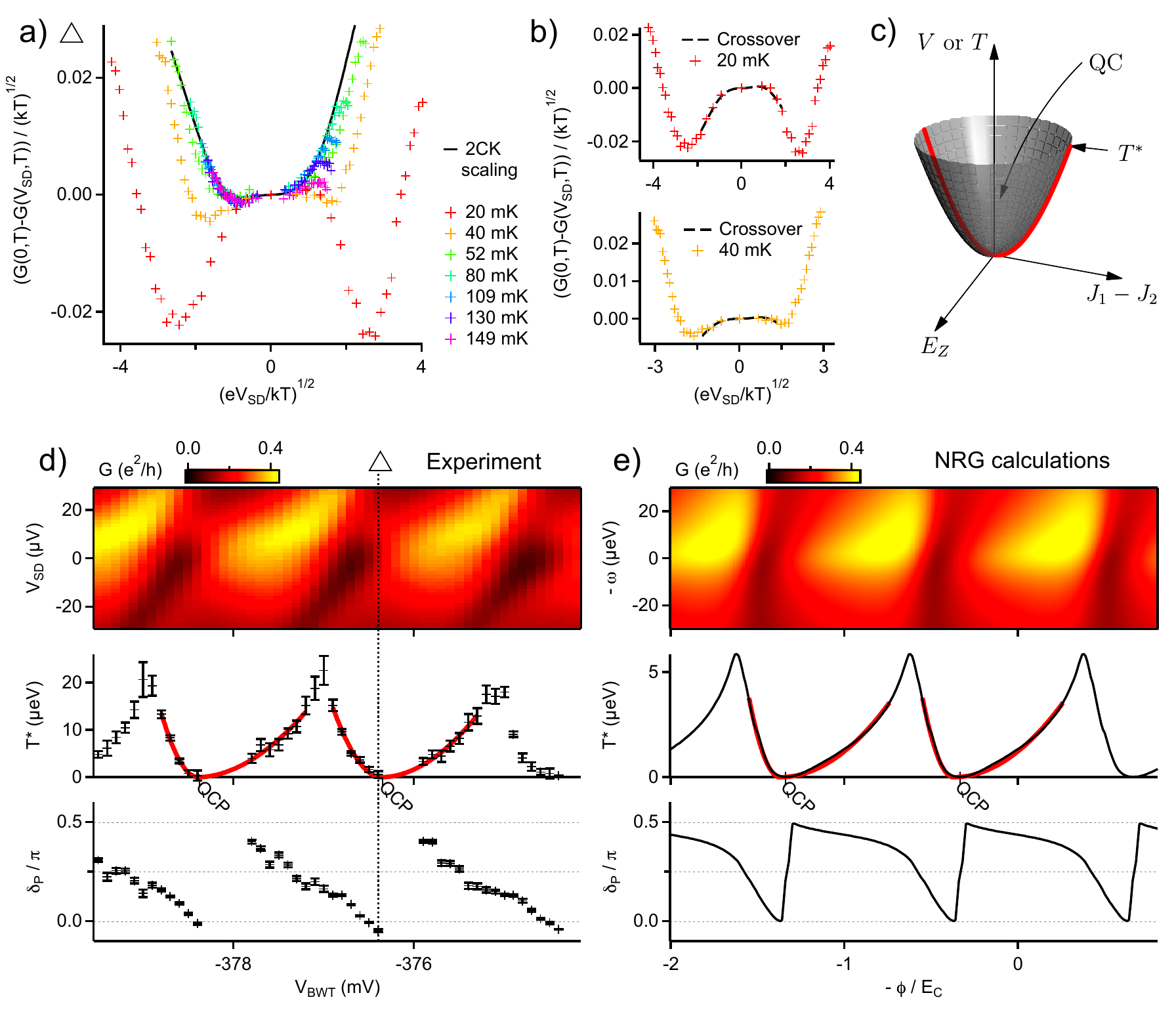}
\caption{\small {\bf Crossover from quantum criticality to a Fermi liquid.} $V_{LP} = -236.0$~mV for experimental data.
{\bf a)} Measured $G(\vsd,T)$. At $V_{BWT} = -376.4$~mV, a thermally broadened spectral function from the 2CK CFT ($\delta_{P} = -0.022\pi$, solid black line) describes the high-$T$ data.
{\bf b)} Same data in a). $G(\vsd)$ at low energies is fit to thermally broadened spectral functions from the crossover theory (top: 20 mK, bottom: 40 mK; $\delta_P = -0.045\pi$, $T^* = 0.5$~$\mu$eV). Fitting details in Methods.
{\bf c)} Quantum criticality occurs for energies above the Fermi liquid scale \ts{} (gray paraboloid), which should depend quadratically on the coupling asymmetry $J_1-J_2$ between the two channels as well as on the Zeeman splitting $E_Z$. We vary $T^*$ by tuning $J_1-J_2$ (cut along red parabola).
{\bf d)} Extraction of \ts{} and $\delta_P$ from measurements. The triangle denotes \vgate{BWT} for a) and b). Top: $G(\vsd, T = 20$~$\rm{mK})$. Middle: \ts{} from crossover theory fits to experimental $G(\vsd,T)$. Red traces are parabolas with $T^* = 0$ at the QCP and unequal scale factors on either side of the QCP. The largest $T^*$ values may not be much less than $T_K$, so the crossover theory is not strictly valid for all \vgate{BWT}. Labels indicate approximate QCP locations. Bottom: $\delta_P$ from the crossover theory fits. Error bars reflect 1 s.d. confidence intervals from the fits.
{\bf e)} Extraction of \ts{} and $\delta_P$ from NRG calculations. Parameters as in \allfig{qptlines}. Top: $G(-\omega)$, rescaled to match maximum $G$ of d). Middle: \ts. Bottom: $\delta_P$.
\label{fig:crossover}
}
\end{figure}

To observe the FL crossover experimentally, we fix \vgate{LP} $= -236$~mV (dashed line in \subfig{qptlines}{c}) and detune the exchange couplings using \vgate{BWT}. Moving slightly away from the QCP so that $T^* \sim T_e$, we still measure a $\sqrt{T}$ scaling collapse for $T > 50$~mK (\subfig{crossover}{a}). These high~$T$ data are fit nicely using the Affleck-Ludwig CFT result with small $\delta_P$ (black line). The clear scaling behavior at high-$T$ can only be observed for $V_{BWT}$ in a small neighborhood around the QCP.  Below 50~mK, prominent deviations from 2CK scaling develop, which we attribute to a crossover into a FL state where the grain screens the dot spin. Near zero bias these low-$T$ traces are fit by the crossover theory with similar, small $\delta_P$ (\subfig{crossover}{b}). We stress this is a non-trivial regime since $T^* \sim T_e$; asymptotics of the FL fixed point are insufficient to describe the observed behavior. For larger $|e\vsd/kT|^{1/2}$, the $|e\vsd|^{1/2}$ dependence of $G(\vsd)$, appearing linear on these axes, heralds a return to 2CK behavior.

Generically, \ts{} should depend quadratically on the strength of symmetry-breaking perturbations near the QCP\cite{Sela2011, Mitchell2012:Universal} (\subfig{crossover}{c}). Measured $G(\vsd,$\vgate{BWT}) reveals periodic zero bias dips that transition sharply to zero bias peaks as \vgate{BWT} is increased (\subfig{crossover}{d}, top). The zero bias dip (peak) corresponds to a $T=0$ ground state where the grain (lead) screens the dot spin; these are separated by a QCP. In \subfig{crossover}{d} (middle), \ts{} depends quadratically on \vgate{BWT} away from the QCP, although the curvature differs between the two sides of the QCP, which have different ground states. This quadratic behavior holds over a larger range of \vgate{BWT} than we might have expected considering that generically the exchange couplings do not depend linearly on gate voltage. The phase shift $\delta_P \sim 0$ on one side of the QCP, and appears to approach $\pi/2$ on the other (\subfig{crossover}{d}, bottom). Between QCPs, $\delta_P$ varies smoothly. $T^*$ and $\delta_P$ are not plotted directly to the right of each QCP, reflecting the ambiguity of fitting a small crossover peak on top of the 2CK peak. Both \ts{} and $\delta_P$ are insensitive to small changes in the range of $\vsd$ used for fitting (supplemental info).

Many features of these observations are corroborated by fitting the crossover theory to spectral functions from NRG, which yield conductance via \refeq{meirwingreen} (Methods). The NRG conductance (\subfig{crossover}{e}, top) shows zero bias dips transitioning into peaks, as well as the shift of the peak toward positive $-\omega$, as in transport spectroscopy (\subfig{crossover}{d}). The $\phi$-dependence of \ts{} (\subfig{crossover}{e}, middle) shows asymmetric parabolas like in the measurements. The extracted $\delta_P$ (\subfig{crossover}{e}, bottom) reproduces the rapid $\pi/2$ phase shift across a QCP, with an otherwise smooth $\phi$-dependence. “The $\pi/2$ shift reflects a sign flip in $S_{\sigma\alpha,\sigma\alpha}$ between distinct FL ground states, where either the grain or leads screen the dot spin \cite{Borda2007}. A perfect correspondence between experiment and NRG should not be expected, since only $U$ and $E_C$ may be extracted directly from measurements. Yet both experiment and NRG are well described by the crossover theory, and key experimental features are reproduced in the NRG calculations.

The experimental and numerical corroboration of analytical results in the vicinity of a QCP is a milestone in our understanding of correlated electron systems, with implications for high-$T_c$ superconductivity and heavy fermions. An essentially identical universal crossover (same $\mathcal{G}\left(\frac{\omega}{T^*},\frac{T}{T^*}\right)$, but different symmetry-breaking perturbations) is expected for the two-impurity Kondo model. Future works will address the full phase diagram of the device, which may host charge 2CK \cite{Matveev1991,Lebanon2003,Anders2004} and SU(4) Kondo regimes~\cite{LeHur2004:MaximizedSpinOrbit,LeHur2007:TransportQDSU4}. Our device geometry could enable Aharonov-Bohm interference measurements to probe phase coherence of low-lying excitations in the non-FL 2CK state~\cite{Borda2007,Carmi2012:TransmissionPhase}, giving new insight into the nature of a local non-FL.

\bibliographystyle{naturemag_ajk}
\bibliography{crossover_BibFile}

\section*{Methods}

\subsection{Measurements \label{sec:meas}}

The measurements are performed in the mixing chamber of a wet dilution refrigerator (Oxford Kelvinox TLM) with a base electron temperature $T_e = 20$~mK, verified by Coulomb blockade thermometry. The device was cooled down with $+300$~mV bias on all gates to enhance charge stability by reducing the range of voltage needed for operation. For all measurements we use an SR830 lock-in amplifier with 1~$\mu$V excitation at 33 Hz and a custom $10^8$~V/A gain current preamplifier (design by H.K. Choi and Y. Chung, see related publication~\cite{Kretinin2012}). A custom voltage source with six 20-bit channels and eight 16-bit channels was used to control the gate and source-drain bias voltages (design by J. MacArthur, assembled and calibrated by AJK).

The biased source lead in any source-drain bias spectroscopy was determined to be weakly coupled to the dot: At zero bias, we pinch off the source lead's coupling to the dot $\Gamma_s$ (e.g. using \vgate{LWT}) and observe a decrease in the overall conductance scale, without appreciable changes in the conductance features after accounting for capacitive shifts from gating. This implies that the unbiased drain lead's coupling to the dot $\Gamma_d$ largely determines the total dot-lead coupling rate, since $\Gamma = \Gamma_s + \Gamma_d \approx \Gamma_d$, i.e. the dot was nearly in equilibrium.  In comparing Fig. 2a and Fig. 2c, the ratio of maximum conductances is 0.464. If these numerical and experimental data are assumed to be directly comparable, then the asymmetry prefactor $4 \Gamma_s \Gamma_d / (\Gamma_s + \Gamma_d)^2 = 0.464$, yielding $\Gamma_s / \Gamma_d = 15\%$.

It is well known that applying source-drain bias will cause unintentional gating as a secondary effect. This would be deleterious to observing quantum critical behavior, which depends sensitively on the dot and grain levels. We compensate for shifts in the grain level by compensating changes in $\vsd$ with changes in \vgate{BWT}. This compensation can be determined easily in the regime $\varepsilon/U > 0$ or $\varepsilon/U < -1$. We expect the grain level to be much more sensitive than the dot level for the same change in energy since $E_C \ll U$.

\subsection{Fitting range \label{sec:range}}

When fitting the crossover theory to experimental data, we fit $G(\vsd,T)$ only in a small window of $\vsd$ of $\pm 6$~$\mu$V around zero, regardless of temperature. A priori, $T^*$ is unknown and it only makes sense to fit $\vsd < \textrm{a few }$\ts. Additionally, thermal broadening of high energy features can in principle spoil the scaling of the low energy features, even for otherwise sensible ranges of $\vsd$. At minimum the 20~and~40~mK traces are used for fitting, but sometimes also the 52~mK and possibly the 70~mK traces, provided $T \lesssim$ \ts{} (the fitting process is somewhat iterative in this respect). Once the temperatures to be used in fitting are decided for a given value of \vgate{BWT}, the fitting considers data from all of those temperatures simultaneously. Fitting the crossover theory to NRG calculations is done analogously (window of $\omega$ of $\pm 6$~$\mu$eV about zero).

\subsection{Relationship of $G(\vsd,T)$ to spectral functions \label{sec:expG}}

The differential conductance $G = dI/d\vsd$ measured from source to drain lead through the small dot is a function of source-drain bias and temperature and can be compared directly to NRG calculations in case of a strongly asymmetrical source-drain coupling. In the case of weak coupling to the biased source electrode ($\Gamma_s \ll \Gamma_d$), the differential conductance can be related to the equilibrium spectral function as

\begin{equation}
  G \left(\vsd,T \right) \approx \frac{2e^2}{h} \frac{4 \Gamma_s \Gamma_d}{\left( \Gamma_s + \Gamma_d \right)^2}  \int_{-\infty}^{+\infty} d\omega
  \left(-\frac{\partial f(\omega-(-e\vsd),T)}{\partial \omega}\right) A (\omega, T).
\label{eq:meirwingreen}
\end{equation}
The asymmetry prefactor is a function of the source and drain couplings, $\Gamma_s$ and $\Gamma_d$, and is assumed to be much smaller than one. Either lead may assume the role of source or drain. The derivative of the Fermi-Dirac distribution $f(\omega,T)$ is convolved with a spectral function $A(\omega,T)$ from the 2CK or crossover descriptions. The spectral function can be related to the \tmatrix{}:

\begin{equation}
A(\omega,T) = - \pi \nu \sum\limits_\sigma \textrm{Im}  \left[ \mathcal{T}_{\sigma \alpha,\sigma \alpha} \left( \omega, T \right) \right] \Big\vert_{ \alpha = -1}\;\;,
\end{equation}
where $\nu$ is the bare density of states in the leads, $\sigma$ is a spin index, and $\alpha$ is a channel index (we fix $\alpha = -1$ for the source and drain leads). The $\mathcal{T}$-matrix represents the scattering between different states induced by the interaction part of the Hamiltonian and can be computed {\it numerically} exactly by NRG. It is related to the quasiparticle self-energy \cite{AltlandSimonsBook}.

\subsection{Fitting expressions for 2CK \label{sec:2ck}}

In equilibrium, the conduction electrons' scattering \tmatrix{}
is proportional to the self-energy.
In case of the quantum dot system considered here,
the latter quantity translates to the Green's function of the d-level of the small dot.
This allows us to use the exact \smatrix{} at the 2CK fixed point \cite{Affleck1993:ExactCFT} and express
the equilibrium spectral function of the small dot in the limit $T^* \ll \omega,T \ll T_K$ as 

\begin{eqnarray} \label{eq:2ckspect}
A(\omega,T) \!\!\! & \approx & \!\! \!
A_{2CK}(\omega,T,\delta_P) = \textrm{Im} \ i\left(1- 3\lambda e^{2i\delta_P} \sqrt{\frac{\pi T}{T_K}} \int\limits_0^1 du \left\{ \vphantom{\frac12} \right. \right. \\
&& \left. \vphantom{\int\limits_0^1} \left.  u^{-i\beta\omega/2\pi} u^{-1/2} (1-u)^{1/2} {}_2 F_1(3/2,3/2;1,u) - \frac{4}{\pi}u^{-1/2}(1-u)^{-3/2} \right\} \right) \nonumber , 
\end{eqnarray}
where ${}_2 F_1(a,b;c,z)$ is the Gauss hypergeometric function,
$\beta$ is inverse temperature, and $\delta_P$ is the scattering phase shift. We fix the dimensionless parameter $\lambda = -0.09$ so that the spectral function drops to half of its $\omega=0$ value at $\omega=T_K$ in the limit $T\rightarrow 0$ \cite{Toth2007,MocaPhysRevB.86.195128}. 
Equation~\eqref{eq:2ckspect} immediately implies that $(A_{2CK}(0, T, \delta_P)-A_{2CK}(\omega, T, \delta_P))\sqrt{T_K/T}$ is a universal function of $\omega/T$, which when convolved with a Fermi function gives the function $Y_{\rm{2CK}}(-e\vsd/kT,\delta_P)$ of equation~\eqref{eq:2ck}. We stress that this $\omega/T$ scaling is a special property of the 2CK fixed point. When fitting the experimental data, we shall assume an asymmetrical coupling to the leads (see Methods sec. \ref{sec:expG}).

\subsection{Fitting expressions for crossover \label{sec:crossover}}

At frequencies and temperatures far below the two-channel Kondo temperature $T_K$, we can use the crossover form of the \tmatrix{} derived in Refs.~\cite{Sela2011,Mitchell2012:Universal} to express the d-level's equilibrium spectral function. Here we obtain the following expression:

\begin{equation}
A(\omega,T)\approx A_{FL}(\omega,T,\delta_P) = \textrm{Im} \  i\left(1-e^{2i\delta_P} \mathcal{G}\left(\tilde{\omega},\tilde{T}\right)\right),
\label{eq:crossoverspect}
\end{equation}
where $\delta_P \approx 0$ ($\delta_P \approx \pi/2$) in case the dot is coupled more strongly to the grain (leads), and

\begin{equation}
\begin{split}
\mathcal{G} \left(\tilde{\omega},\tilde{T} \right)&=\frac{ \frac{-i}{\sqrt{2 \pi^3 \tilde{T}}}}{\tanh \frac{\tilde{\omega}}{2 \tilde{T}}} \frac{\Gamma \left(\frac{1}{2}+\frac{1}{2 \pi \tilde{T}} \right)}{\Gamma \left(1+\frac{1}{2 \pi \tilde{T}} \right)}\times  \\
&\int_{-\infty}^\infty dx \frac{ e^{ ix \tilde{\omega}/\pi \tilde{T}}}{\sinh x} {\rm{Re}} \left[ _{2}F_1\left(\frac{1}{2},\frac{1}{2};1+\frac{1}{2 \pi \tilde{T}},\frac{1-\coth x}{2} \right) \right]\; ,
\end{split}
\label{eq:G}
\end{equation}
is a universal function of rescaled energy $\tilde{\omega} = \omega/T^*$ and temperature $\tilde{T} = T/T^*$. For equation~\eqref{eq:G} only, $\Gamma$ is the gamma function, not a tunnel rate. Again, when fitting to experimental data, the spectral function must be thermally broadened (see Methods sec. \ref{sec:expG}).

\section*{Acknowledgments}

We are grateful to S. Amasha, Y. Oreg, A. Carmi, E. Sela, A. K. Mitchell, and M. Heiblum for discussions; H. K. Choi, Y. Chung, and J. MacArthur for electronics expertise; M. Heiblum for use of his lab during initial device characterization; H. Inoue, N. Ofek, O. Raslin, and E. Weisz for fabrication guidance; F. B. Anders, E. Lebanon, and the late A. Schiller for their calculations which guided prior experimental work; M. Stopa for his SETE software for electrostatic quantum dot modeling. The device fabrication was done in the Braun Submicron Center at Weizmann Institute of Science, with final fabrication steps done at Stanford Nano Shared Facilities (SNSF) at Stanford University. This work was supported by the Gordon and Betty Moore Foundation through Grant GBMF3429, the Hungarian research grant OTKA K105149, the Poland National Science Center Project No. DEC-2013/10/E/ST3/00213, the EU Grant No. CIG-303 689, the National Science Foundation Grant No. DMR-0906062, and U.S.-Israel BSF Grant No. 2008149. AJK and LP have been supported by a Stanford Graduate Fellowship. SETE calculations were run on the Odyssey cluster supported by the FAS Division of Science, Research Computing Group at Harvard University. NRG calculations were performed at Pozna\'{n} Supercomputing and Networking Center.

\section*{Author contributions}

AJK, GZ, and DGG designed the experiment. AJK and LP performed the measurements. IW, CPM, and GZ performed the NRG calculations. CPM and IW contributed equally to the theoretical analysis. AJK, LP, CPM, IW, GZ, and DGG analyzed the data. AJK designed and fabricated the devices, with e-beam lithography from DM, using heterostructures grown by VU. AJK and LP wrote the paper with critical review provided by all other authors.

\section*{Competing financial interests}
The authors declare no competing financial interests.

\clearpage


\section*{Supplementary material}

# Supplemental information for "Universal Fermi liquid crossover and quantum criticality in a mesoscopic system"

A. J. Keller[1], L. Peeters[1], C. P. Moca[2,3], I. Weymann[4], D. Mahalu[5], V. Umansky[5], G. Zaránd[2], and D. Goldhaber-Gordon[1,*]

[1]Geballe Laboratory for Advanced Materials, Stanford University, Stanford, CA 94305, USA
[2]BME-MTA Exotic Quantum Phases "Lendület" Group, Institute of Physics, Budapest University of Technology and Economics, H-1521 Budapest, Hungary
[3]Department of Physics, University of Oradea, 410087, Romania
[4]Faculty of Physics, Adam Mickiewicz University, Poznań, Poland
[5]Department of Condensed Matter Physics, Weizmann Institute of Science, Rehovot 96100, Israel
[*]Corresponding author; goldhaber-gordon@stanford.edu

# Contents

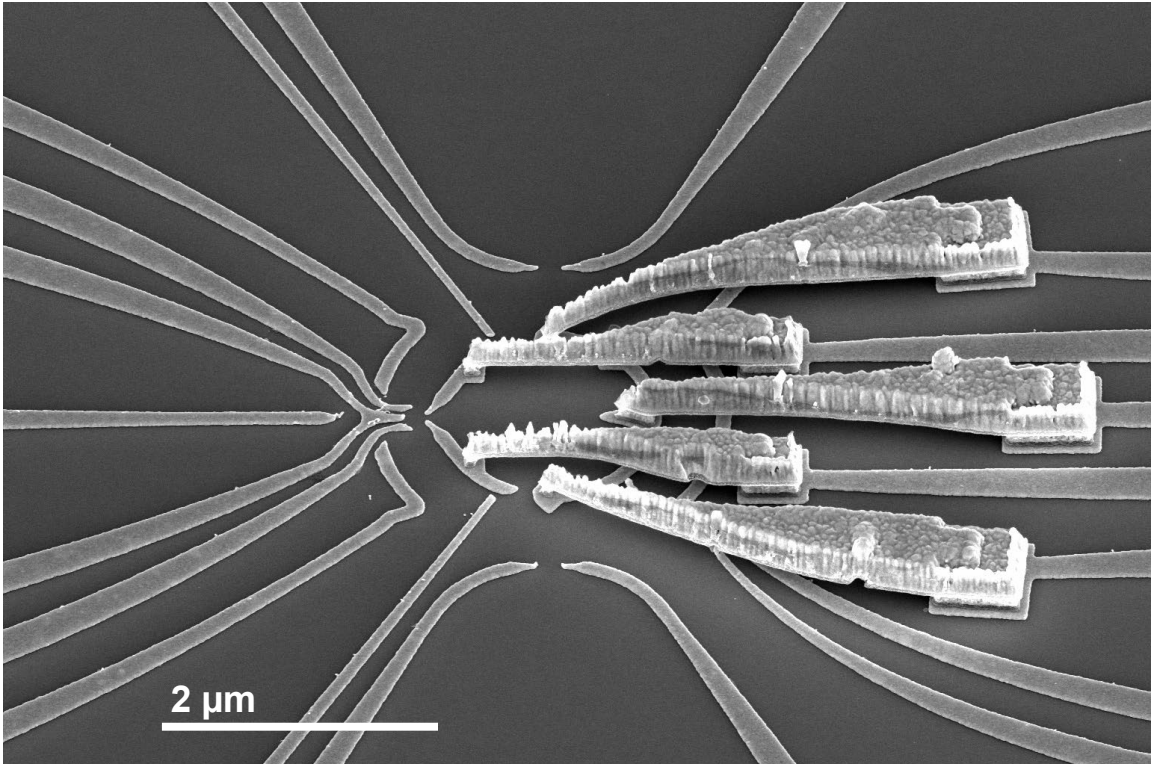

Figure S1: SEM micrograph of a nominally identical device (5 kV acceleration voltage). The device is tilted 40° with respect to normal incidence.

## S1 Device

The 2DEG is 50 nm deep and has an electron density $n = 3.3 \times 10^{11}$ cm$^{-2}$ and mobility $\mu = 1.2 \times 10^6$ cm$^2$/Vs.

Figure 1a shows a top-down SEM micrograph of the device. This view is appropriate for labeling the gate electrodes and explaining the function of each gate, but the air bridges are hard to see. In Fig. S1 we show a view of the device at a 40° tilt with respect to normal incidence. The five air bridges clearly rise above the gate electrodes underneath. The device is rotated approximately 180° with respect to the orientation of Fig. 1a.

As initially fabricated, the bridges did not make good electrical contact to the gates. This problem was remedied with an in-situ platinum deposition procedure to be described in a forthcoming publication.[1]

# S2 Hamiltonian

In our numerical calculations the dot-grain system is modeled by the following Hamiltonian:

$$H_{\text{device}} = H_{\text{dot}} + H_{\text{grain}} + H_{\text{leads}} + H_{\text{tunneling}}, \tag{1}$$

where

$$H_{\text{dot}} = \sum_\sigma \varepsilon\, d_\sigma^\dagger d_\sigma + U \hat{n}_\uparrow \hat{n}_\downarrow, \tag{2}$$

describes the dot, with $\varepsilon$ the on-site energy, $d_\sigma^\dagger$ the creation operator of an electron with spin $\sigma$ and $\hat{n}_\sigma = d_\sigma^\dagger d_\sigma$ representing the occupation number operator. $U$ is the correlation energy between the two electrons residing in the dot. The grain is described by

$$H_{\text{grain}} = \sum_{p,\sigma} \varepsilon_p a_{p\sigma}^\dagger a_{p\sigma} + \frac{E_C}{2}(\hat{n}_g - N_0)^2 + \phi(\hat{n}_g - N_0). \tag{3}$$

The creation operator of a spin-$\sigma$ electron with momentum $p$ and energy $\varepsilon_p$ in the grain is denoted by $a_{p\sigma}^\dagger$, $E_C$ is the charging energy of the grain, while $-\phi$ plays the role of a gate voltage. $\hat{n}_g$ is the electron number operator of the grain, $\hat{n}_g = \sum_{p,\sigma} :a_{p\sigma}^\dagger a_{p\sigma}:$, and $N_0$ denotes the number of excess electrons in the electrically neutral grain ($\phi = 0$). The non-interacting quasiparticles in the leads are described by:

$$H_{\text{leads}} = \sum_{\alpha=\{U,L\}} \sum_{k,\sigma} \varepsilon_{\alpha k} c_{\alpha k \sigma}^\dagger c_{\alpha k \sigma}, \tag{4}$$

where $c_{\alpha k \sigma}^\dagger$ is the creation operator of a spin-$\sigma$ electron with momentum $k$ and energy $\varepsilon_{\alpha k}$ in the upper ($\alpha = U$) or lower ($\alpha = L$) lead. The last term in (1) is the tunneling Hamiltonian, which is given by

$$H_{\text{tunneling}} = t_G \sum_{p,\sigma} \left(a_{p\sigma}^\dagger d_\sigma + d_\sigma^\dagger a_{p\sigma}\right) + \sum_{\alpha=\{U,L\}} \sum_{k,\sigma} t_\alpha \left(c_{\alpha k \sigma}^\dagger d_\sigma + d_\sigma^\dagger c_{\alpha k \sigma}\right). \tag{5}$$

The tunnel matrix elements between the leads (grain) and the dot are denoted by $t_\alpha$ ($t_G$) and are assumed to be independent of momentum. The strengths of the couplings are given by $\Gamma_G = \pi \nu_G |t_G|^2$ and $\Gamma_\alpha = \pi \nu_\alpha |t_\alpha|^2$, respectively, where $\nu_\alpha (\nu_G)$ is the density of states for lead $\alpha$ (grain). In the numerical renormalization group (NRG) calculations the energy spectrum of the grain is assumed to be continuous and the densities of states for leads and grain are taken to be constant and equal: $\nu_\alpha = \nu_G = \nu = 1/(2D)$, with $D \equiv 1$ being the

band halfwidth used as the energy unit in NRG calculations.

In Hamiltonian (1) we have neglected the dot-grain capacitive coupling, which can give rise to a term of the form, $U_{dg}(\hat{n}_g - N_0)\hat{n}_d$, where $\hat{n}_d = \hat{n}_\uparrow + \hat{n}_\downarrow$. An estimate for $U_{dg}$ is extracted experimentally in Section S4, and is believed to play no role for the present analysis.

## S3 Summary of the NRG calculations

### S3.1 NRG calculations

To solve the Hamiltonian (1) we use the numerical renormalization group method.[2,3] First, we introduce the collective charge operators (bosonic operators) for the grain,[4,5]

$$\hat{N} = \sum_{m=-\infty}^{\infty} m|m\rangle\langle m| \quad \text{and} \quad \hat{N}^{\pm} = \sum_{m=-\infty}^{\infty} |m \pm 1\rangle\langle m|. \tag{6}$$

Strictly speaking, the identity, $\hat{N} = \hat{n}_g$, must be fulfilled, but within the NRG approach this constraint can be relaxed by treating $\hat{N}$ as an independent quantity. This is possible as the spectral properties of the system are not sensitive to the exact number of conduction electrons present in the grain in the limit of infinitely small level spacing. To extract the finite size spectrum and determine the location of the two-channel Kondo (2CK) lines, however, a projection to the physical subspace was necessary. In our calculations we took into account seven charges in the grain. Using the above charge operators, the grain part of the Hamiltonian (3) can be rewritten as

$$H_{\text{grain}} = \sum_{p,\sigma} \varepsilon_p a_{p\sigma}^\dagger a_{p\sigma} + \frac{E_C}{2}(\hat{N} - N_0)^2 + \phi(\hat{N} - N_0). \tag{7}$$

The $\hat{N}^{\pm}$ operators capture the charging transitions of the grain and enter explicitly in the tunneling Hamiltonian (5), which now reads

$$H_{\text{tunneling}} = \sqrt{\frac{2\Gamma_G}{\pi}} \sum_{p,\sigma} \left(\hat{N}^+ a_{p\sigma}^\dagger d_\sigma + d_\sigma^\dagger a_{p\sigma} \hat{N}^-\right) + \sqrt{\frac{2\Gamma}{\pi}} \sum_{k,\sigma} \left(c_{k\sigma}^\dagger d_\sigma + d_\sigma^\dagger c_{k\sigma}\right). \tag{8}$$

The first term in (8) describes the dot-grain tunneling, while the second term accounts for the dot-leads coupling. This second term is obtained by performing an orthogonal transformation[6] from the two-lead basis to an effective single lead with resultant coupling $\Gamma = \Gamma_L + \Gamma_U$.

The resulting Hamiltonian consists then of two conduction bands coupled to a complex impurity composed of the grain ($\hat{N}$) and dot degrees of freedom.

The core of the NRG procedure is the logarithmic discretization of the conduction band with discretization parameter $\Lambda$ and mapping of the conduction band onto a semi-infinite chain with exponentially decreasing hoppings. The Hamiltonian can then be diagonalized in an iterative fashion. In our calculations we used discretization parameter $\Lambda = 2$ and kept 4000 states at each iteration. We also exploited the $SU(2)$ symmetry of the total spin and two $U(1)$ symmetries for $\hat{N}_1 = \hat{n}_d + \hat{n}_{cb} + \hat{n}_g$ and $\hat{N}_2 = \hat{n}_g - \hat{N}$, where $\hat{n}_{cb}$ is the electron number operator in the first conduction channel (leads coupled to the dot) and $\hat{n}_g$ is the electron number operator in the second channel (the grain). We performed the full density-matrix numerical renormalization group calculations (fDM-NRG),[2,7–9] employing the Budapest Flexible DM-NRG code,[3] to compute the normalized dimensionless spectral function, $A(\omega, T) \equiv \pi(\Gamma_L + \Gamma_U)A_d(\omega, T)$, where $A_d(\omega, T)$ is the spectral function for the $d^\dagger_\sigma$ operators that describe the dot level. The linear conductance through the small dot can be then determined with the equation

$$G = \frac{2e^2}{h} \frac{4\Gamma_U \Gamma_L}{(\Gamma_U + \Gamma_L)^2} \int d\omega \left( -\frac{\partial f(\omega, T)}{\partial \omega} \right) A(\omega, T) , \qquad (9)$$

where $f(\omega, T)$ is the Fermi-Dirac distribution function.

## S3.2 Shifting of NRG calculations in Fig. 2b

In Fig. 2b we incorporate a linear $\varepsilon$-dependent shift into $\phi$ to obtain agreement between NRG calculations and experiment (Fig. 2c). The agreement is obtained by first rescaling the NRG calculations so that the maximum value of $G$ is the same. The global scaling takes into account the source-drain coupling asymmetry, as explained in Methods. Then, the sharp features in the cut taken at $V_{\mathrm{LP}} = $ -260 mV are compared with NRG calculations to establish $V_{\mathrm{LP}} = -260$ mV $\sim -\varepsilon/U = 0.55$. Finally, another cut for fixed $V_{\mathrm{LP}}$ is taken to establish a linear relationship between $V_{\mathrm{LP}}$ and $-\varepsilon/U$. The two points give a linear relationship between $V_{\mathrm{LP}}$ and $-\varepsilon/U$. One global offset in $-\phi/E_C$ then suffices to give good agreement everywhere. Using this method we find $-\Phi/E_C = -\phi/E_C - 3.1(-\varepsilon/U - 1.5)$, where $-\Phi/E_C$ is the vertical axis of Fig. 2b. Physically, the linear dependence of $\Phi$ on $-\varepsilon/U$ can be understood as a consequence of the indirect capacitive coupling between $V_{\mathrm{LP}}$ and the grain.

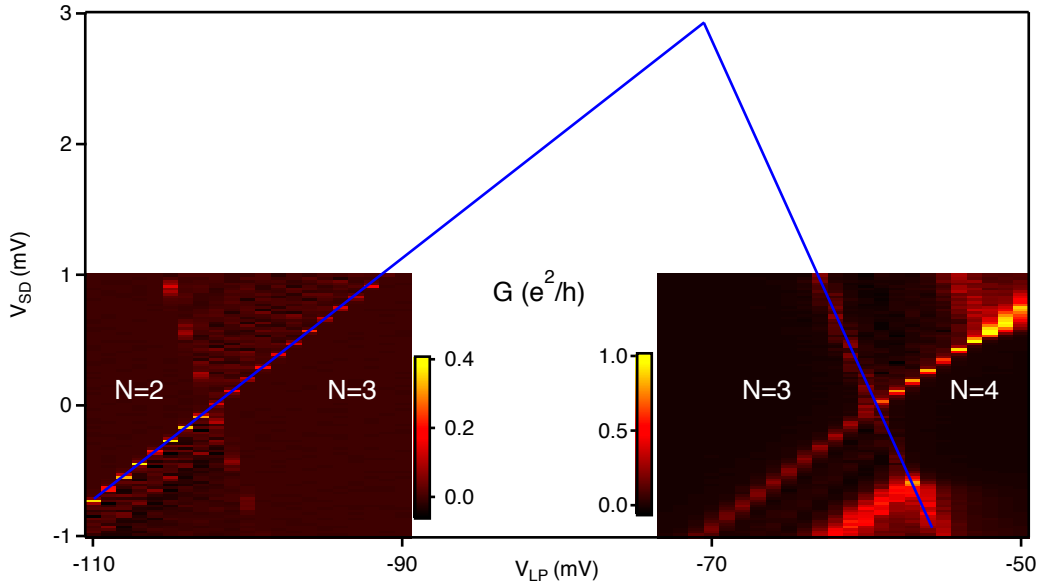

Figure S2: Measurement of $U$. Within each Coulomb diamond we label the number of electrons on the dot as determined by charge sensing techniques. The intersections of the lines indicate $U \sim 2.9$ meV.

## S4   Extracting device parameters

From measurements in the Coulomb blockade regime we are able to determine the dot charging energy $U$ and the grain charging energy $E_C$. In determining $E_C$ we justify treatment of the grain as a continuum. We also determine bounds on any dot-grain charging energy $U_{dg}$ neglected in the model.

The dot charging energy $U = 2.9$ meV is determined from source-drain bias spectroscopy of the dot (Fig. S2). In previous cooldowns $U$ has varied between 1 and 3 meV, perhaps owing to how $U$ depends sensitively on the number of electrons in the few electron regime. We use $U = 2$ meV as the model parameter in NRG calculations, and note that the calculations should be relatively insensitive when $U > D = 1$ meV, the electronic half bandwidth used in calculations. This value of $D$ corresponds roughly to the internal level spacing on the small dot, providing a high energy cut-off.

The grain charging energy $E_C = e^2/C = 160$ $\mu$eV is measured by source-drain bias spectroscopy of the grain (Fig. S3). We compare this measurement to geometric estimates. A common rule of thumb is that upon gate depletion, the extent of the depletion region extends as far from the gate laterally as the 2DEG is deep. This means that the area of the grain should be the area outlined by the gates, less some area in a $\sim 50$ nm dead region

6

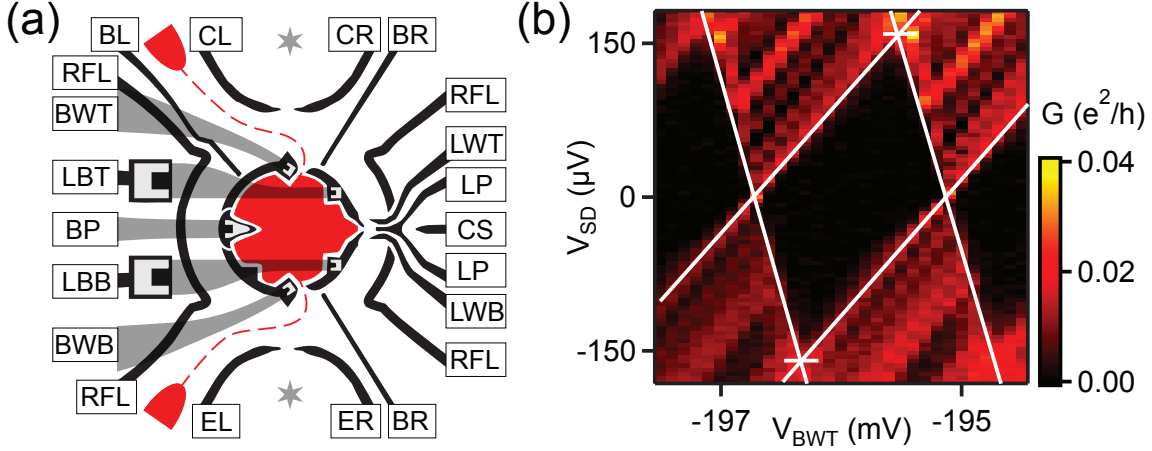

Figure S3: Measurement of $E_C$. a) Measurement scheme. $G(V_{sd})$ is measured using the grain's own pair of measurement leads, which are isolated from the measurement leads of the dot by depleting gate BR. Gate BL is depleted to avoid shorting conductance through the channel just left of the grain. b) $G(V_{sd})$ through the grain in the Coulomb blockade regime ($T = 20$ mK). The intersections of the lines indicate $E_C \sim 160$ μeV.

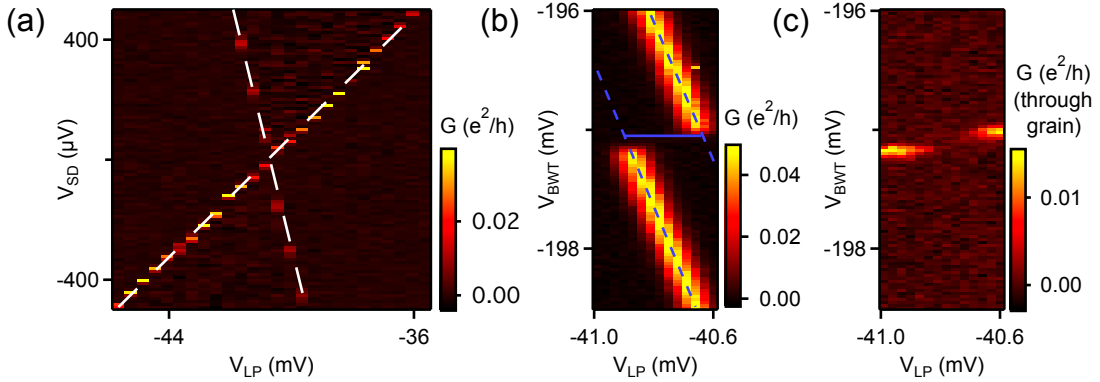

Figure S4: Bounding $U_{dg}$ from measurements in the Coulomb blockade regime. (a) $G(V_{sd}, V_{LP})$ through the dot in the Coulomb blockade regime, with both the dot and grain formed. Here $V_{BWT}$ is such that the grain is Coulomb blockaded. From the slopes of the dashed lines overlaid on the peaks in the data, we determine lever arms $\alpha_{LP}$ = 0.081 and $V_{sd}$ = 0.194. (b) $G(V_{BWT}, V_{LP})$ through the dot at zero $V_{sd}$. Peaks in $G$ correspond to Coulomb blockade on the dot being lifted; the splitting implies finite $U_{dg}$. For fixed $V_{BWT}$ the difference in peak positions gives the dot-grain charging energy $U_{dg} = (e)(\alpha_{LP})(\Delta V_{LP}) = 0.081 * 0.26$ meV $= 21$ μeV. Dot-grain tunneling is negligible in this limit. (c) Conductance through the grain appears where expected given the interpretation of (b). The conductance is measured with gates BL and BR depleted, measuring through the two point contacts formed by gate pairs LBB / BWB and LBT / BWT.

7

adjacent to the gates. The red shaded area in Fig. S3a is 1.2 $\mu$m$^2$. Approximating the capacitance of the grain as that of a flat disk with radius $r$, $C = 8\epsilon r$ where $\epsilon = 13\epsilon_0$ for GaAs and the effective $r = 0.62$ $\mu$m. This gives an expected $E_C = 280$ $\mu$eV, which is within a factor of two of the measurement. In a previous cooldown of the same device we measured $E_C = 150$ $\mu$eV, which we use as the model parameter in NRG calculations.

In designing the device we aimed for as large an $E_C$ as possible while still being able to imagine a near continuum of states in the grain. The level spacing may be estimated by considering a particle in a 2D box. The level spacing $\Delta = \hbar^2 \pi^2 / 2mA$, where $A$ is the area of the box and $m = 0.067 m_e$ is the effective mass in GaAs. Using the design area $A = 1.2$ $\mu$m$^2$ we find $\Delta = 4.6$ $\mu$eV $= 2.6 k T_e$, where $T_e = 20$ mK. If we instead take $A = 0.93$ $\mu$m$^2$ inferred from measurement of $E_C$ and the approximation for the capacitance, we find $\Delta = 6.0$ $\mu$eV$= 3.4 k T_e$. In either case, $\Delta$ is no more than factors of a few times $T_e$, keeping in mind that the width of the Fermi-Dirac distribution is approximately $3.5 k T_e$. This implies that the grain is indeed acting as a metallic grain at all measured temperatures. In Fig. S3, it appears that the typical level spacing (spacing in $V_{sd}$ between diagonal lines) is larger than anticipated. The peak conductance can differ significantly for each level, which reflects a distribution in source-drain coupling asymmetry from level to level. Some levels may not be visible if their source-drain coupling asymmetry is strong.

Measurements of $G(V_{BWT}, V_{LP})$ yield $U_{dg} = 21$ $\mu$eV (Fig. S4). This analysis considers the dot-grain system as a capacitively-coupled double quantum dot. The $U_{dg}$ we extract should be thought of as an upper bound—the gate voltages are set to a very different regime where $\Gamma_G$ is negligible, unlike in the paper. When tuning between this regime and the regime where $\Gamma_G \sim \Gamma$, it appears as if the splitting of the lines in Fig. S4 goes to zero long before $\Gamma_G$ becomes a significant fraction of $\Gamma$, perhaps implying that $U_{dg} \to 0$.

## S5   Scaling along the quantum critical lines

In Fig. 2e we demonstrate that measured $G(V_{sd}, T)$ falls onto a scaling curve derived from the conformal field theory (CFT) results of Affleck and Ludwig.[10] In Fig. S5 we demonstrate the scaling at other points along the 2CK lines. In most examples, the 2CK scaling behavior is faithfully reproduced by the data except perhaps at $T = 20$ mK.

A priori, the deviations could be attributed to finite $T^*$, perhaps because data were not taken finely enough. However, it seems that the deviations typically appear most pronounced on the positive-$V_{sd}$ side. Another possibility could be that true zero $V_{sd}$ drifted slightly over

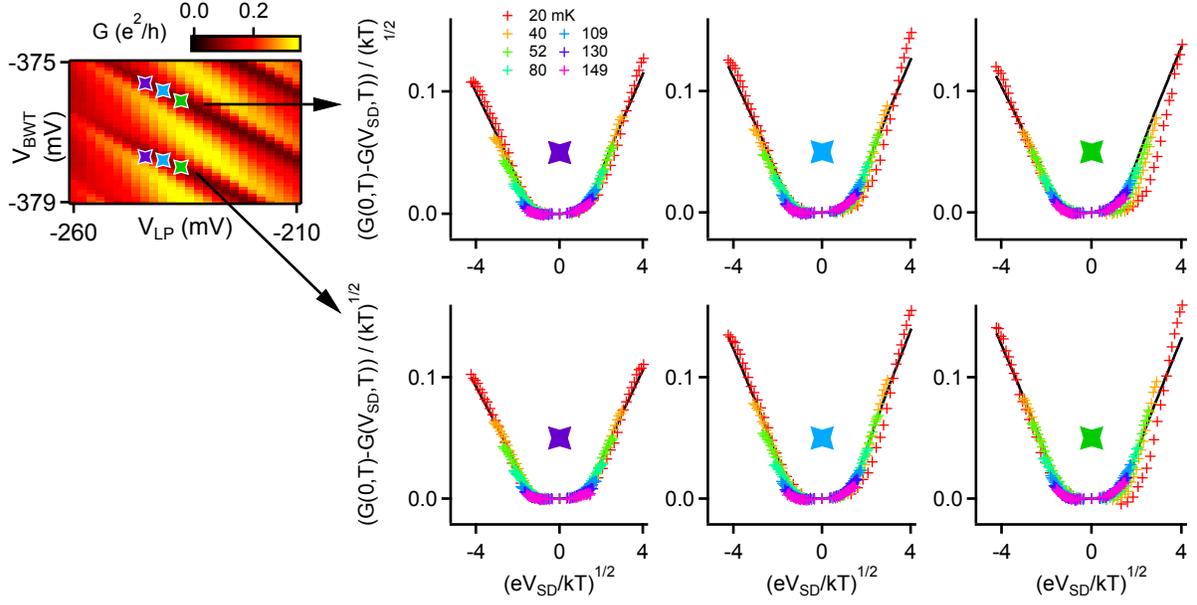

Figure S5: Top-left: Measured $G(V_{\text{LP}}, V_{\text{BWT}})$ of Fig. 2c. Panels at right: Measured $G(V_{sd}, T)$ at six points on 2CK lines in the $(V_{\text{LP}}, V_{\text{BWT}})$ plane. Black lines are fits to thermally broadened spectral functions from Affleck and Ludwig (Ref.[10]) with small phase shifts from potential scattering.

the course of the measurement. Typically some small applied $V_{\text{sd}}$ is required to compensate for an offset voltage at the current amplifier input, but at base temperature the quality of the scaling collapse is sensitive to errors of just 1 $\mu$V in identification of true zero $V_{\text{sd}}$.

## S6  Sensitivity of $T^*$ and $\delta_P$ to fitting range

In Fig. 3 we use the crossover CFT to fit experimental data and thereby extract the Fermi liquid scale $T^*$ and the scattering phase shift $\delta_P$. The fitting procedure uses a limited range of $V_{\text{sd}}$ ($\pm 6$ $\mu$V) and it is argued that this is a conservative approach.

In Fig. S6 we show that the fitting is insensitive to small changes in the fitting range. At each value of $V_{\text{BWT}}$, we try and extract $T^*$ and $\delta_P$ for nine different ranges of bias voltage, which we obtain by starting with (-6, +6 $\mu$V) and adding or subtracting a point on either end, e.g.: (-7.5, +6), (-6, +6), (-6, +7.5), (-4.5, +6), etc. It is important not to add so many points that data outside the validity of the theory are included. However, subtracting too many points may degrade the fit quality.

After finding $T^*$ and the error in $T^*$ reported by the fits, we consider all nine fitting ranges to give independent estimates of $T^*$, and find the weighted mean $T^*$, weighted by the

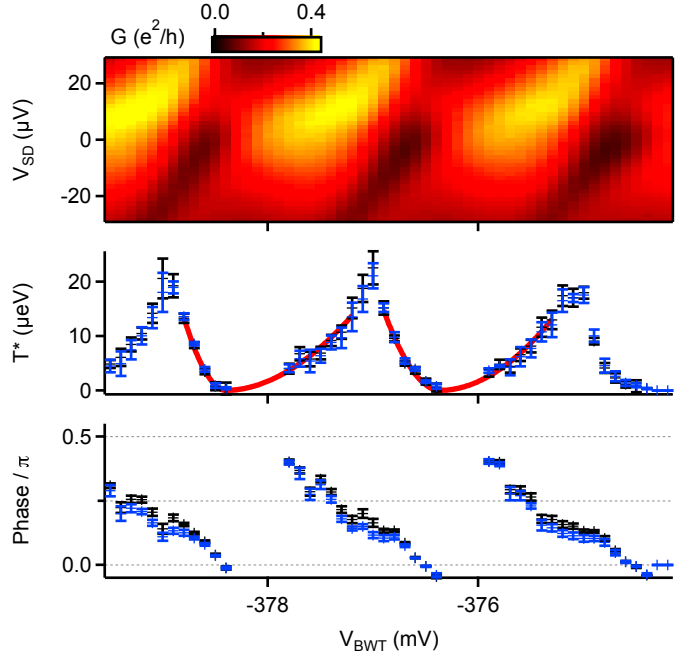

Figure S6: Sensitivity of $T^*$ and $\delta_P$ to fitting range. Top panel: source-drain bias spectroscopy at $T = 20$ mK exactly as in Fig. 3. Middle and bottom panels: The black points and red curves in the $T^*$ and $\delta_P$ panels are exactly as in Fig. 3. The blue points correspond to the weighted mean of extracted $T^*$ and $\delta_P$ for an ensemble of fitting ranges. The error bars on the blue points correspond to the standard deviation of the weighted mean.

10

errors from each fit. The error bars show the standard deviation of the weighted mean, and indicate the spread of $T^*$ values returned by the fits. We do the same for $\delta_P$. Varying the fitting range by small amounts does not seem to contribute significantly to the uncertainty in $T^*$ and $\delta_P$.


\end{document}